\documentclass[preprint,12pt]{aastex}
\usepackage{emulateapj5}
\usepackage{graphicx}
\usepackage{rotate}
\usepackage{amsmath}
\usepackage{amssymb}

\begin{document}

\title{On the role of magnetic fields in abundance determinations}

\author{J.M.~Borrero}
\affil{High Altitude Observatory (NCAR), 3080 Center Green Dr. CG-1, Boulder, CO 80301, USA}
\email{borrero@ucar.edu}

\begin{abstract}
{Although there is considerable evidence supporting
an ubiquitous magnetic field in solar/stellar photospheres, its impact 
in the determination of abundances has never been quantified. In this work 
we investigate whether the magnetic field plays a measurable role for this kind 
of studies. To that end, we carry out simulations of spectral line formation in the 
presence of a magnetic field, and then use those profiles to
derive the abundance of several atomic species (Fe, Si, C and O) neglecting the magnetic field.
In this way, we find that the derived iron abundance can be significantly biased, 
with systematic errors up to 0.1 dex. In the case of
 silicon, carbon and oxygen their role is very marginal (errors smaller than 0.02 dex). We 
also find that the effect of the magnetic field strongly depends on its 
inclination with respect to the observer. We show that fields that are aligned
with the observer lead to an underestimation of the real abundance, whereas more inclined ones overestimate it. 
In the case of a mixture of fields with different inclinations these effects 
are likely to partly cancel each other out, making the role of the magnetic field even less important.
Finally, we derive a simple model that can be used to determine the suitability
of a spectral line when we wish to avoid the bias introduced by the neglect of the magnetic field.}
\end{abstract}

\shorttitle{Magnetic field and abundance determinations}
\shortauthors{J.M.~Borrero}
\maketitle

\section{Introduction}

From the very first determinations of solar and stellar abundances
using 1D semi-empirical atmospheric
models (e.g. Lambert 1968, Lambert \& Warner, B. 1968, Grevesse 1968, Garz et al. 1969)
to more recent values obtained from state-of-the-art 3D hydrodynamical
simulations (e.g. Asplund et al. 2005a), the role
of the magnetic field has rarely been considered.
This would be strictly valid only if atomic transitions with zero Land\'e factors
are used in the analysis. Unfortunately this has never been the case.
The reason for this is that there are few magnetic insensitive spectral 
lines having accurate oscillator strengths. In the case of the Sun, the role
of the magnetic field has been avoided by arguing that the FTS-disk 
center spectral atlas (Brault \& Neckel 1987; Neckel 1999), which is the
most common source to compare observed and simulated line profiles, 
was recorded around a quiet Sun region. 

The existence of significant magnetic flux in quiet Sun regions
has passed unnoticed because this magnetic field organizes in patches 
of opposite polarity\footnote{In spectropolarimetry, the term
positive polarity refers to magnetic fields pointing towards the observer: 
inclination angles $\gamma < 90^{\circ}$, while negative polarities refer
to magnetic fields pointing away from the observer: $\gamma > 90^{\circ}$} 
over very small scales, leading to a cancellation 
of the polarization signals.  However, there is now strong evidence 
supporting the omnipresence of magnetic fields in regions previously
thought to be void of them (Trujillo Bueno et al. 2004; 
Manso Sainz et al. 2004). The details about its actual strength 
and distribution are subject to debate, however.

This is even more critical when disk integrated data is used in 
stellar abundances studies (e.g. Allende Prieto et al. 2002), in particular if 
the star is magnetically active (Ap-Bp types), as regions of strong
magnetic field (i.e.: starspots) can have a large influence.

Although the magnetic field will mainly affect the polarization signals, 
it also has an impact on the intensity profiles (Stokes $I$), being 
this particularly important because this effect on the intensity profiles 
adds up regardless of its polarity. 
If not accounted for, this might lead to systematic errors 
in the abundance values derived from the fitting. 
The goal of this work is to asses the importance of neglecting the
role of magnetic fields. To that end we
will perform simplistic simulations of intensity profiles of various important
atomic elements in the presence of a magnetic field. In Section 2 we describe
the synthesis code employed, the spectral lines used and we briefly review
the Zeemann effect applied to Stokes $I$. In Section 3 we study how the strength and inclination of the
magnetic field taintes the inferred abundances of iron, silicon
carbon and oxygen when the existence of this field is neglected. In Section 4 we
derive a simple model that is able to quantitatively predict
the commited error. Section 5 is devoted to studying if results from
our simple modeling could differ significantly if the same investigation
is performed using more realistic 3D MHD models. Section 6 summarizes
our findings and anticipates possible future work.

\section{Spectral lines synthesis}
\subsection{Synthesis code}

We have employed the SIR code (Ruiz Cobo \& del Toro Iniesta 1992) to
produce synthetic spectral lines. This code solves the radiative transfer equation in the presence of
a magnetized plasma. Although SIR synthesizes the full Stokes vector, we will 
restrict ourselves to consider only the total intensity, Stokes $I$, as done in most 
abundance studies. In addition, SIR allows for the magnetic field vector to be a function
of the optical depth, but in this work we will consider it constant.

The Harvard-Smithsonian Reference Atmosphere (HSRA; Gingerich et al. 1971) 
was used in our calculations. Note that using this 1D LTE semi-empirical model
means that we study the impact of the magnetic field alone, that is, 
we assume that other important ingredients of the spectral line formation are 
already being accounted for. For instance, the magnetic field couples with
the energy and momentum equation, resulting in a modification
of the temperature stratification. In our analysis we are assuming that the correct temperature
stratification and convective velocity fields are known (i.e.: obtained through realistic 3D simulations)
and therefore we focus on the magnetic field. If this was not the case, small errors in the 
temperature or velocities would dominate over the neglect of the magnetic field.

To account for the convective broadening of the spectral lines we
use, unless otherwise specified, a macroturbulent velocity of 2 km s$^{-1}$ and a microturbulent velocity of 1 km s$^{-1}$.
We anticipate (see Section 5) that these values have no particular consequences in our discussion.

\subsection{Spectral line selection}

We have decided to focus our study on four important atomic elements: \ion{Fe}{1},
\ion{Si}{1}, \ion{C}{1} and \ion{O}{1}. Iron is particularly important because it shows a large 
number of atomic transitions at visible wavelengths, and therefore it is often used to
investigate solar and stellar atmospheres. In addition, it is commonly used to distinguish whether a
star is first or second generation since heavier elements are only produced in Supernova 
explosions (Christlieb et al. 2002). Silicon is important in the solar context as it is used, 
together with iron, to compare with meteoritic abundances (Asplund 2000). Oxygen
and carbon's importance comes from being the third and fourth most abundant elements in the Universe,
respectively. In addition, they have both been targeted as being responsible for the
lowering of the solar metallicity (Asplund et al. 2005a), that has caused
major discrepancies between helioseismic inversions and solar models (Castro et al. 2007).
Therefore, it is worthwhile investigating whether magnetic fields could have 
something to say in this regard.

In total we consider 57 spectral lines: 29 of \ion{Fe}{1}, 15 of \ion{Si}{1}, 9 of \ion{C}{1} 
and 4 of \ion{O}{1}.  Their properties are summarized in Table 1. They have been adopted from Asplund (2000) and 
Asplund et al. (2000, 2004, 2005b). We have rejected those lines for which accurate 
collisional parameters (under the ABO theory; Anstee \& O'Mara 1995; Barklem \& O'Mara 1997; 
Barklem et al. 1998) were not found. Note that only one line, \ion{Si}{1} 5665.555 \AA\ has
a zero Land\'e factor: $g_{\rm eff}=0$. Several neutral iron and carbon lines are potentially
very sensitive to magnetic fields: $g_{\rm eff} \gtrapprox 2$.

\subsection{Intensity profiles under the presence of a magnetic field}

Ignoring off-diagonal elements in the propagation matrix, the main contributor to the intensity
profiles is $\eta_I$ (Wittmann 1974; del Toro Iniesta 2003):

\begin{equation}
\eta_I = 1 + \frac{\eta_0}{2}\left\{\phi_p \sin^2\gamma + \frac{1}{2}[\phi_b+\phi_r](1+\cos^2\gamma)\right\}
\end{equation} 

where $\eta_0$ is related to the abundance of the element, the excitation potential of the lower level
and the transition probability of the atomic transition (Landi Degl'Innocenti 1976). $\gamma$ refers 
to the inclination of the magnetic field vector with respect to the observer. Since it appears
as $\cos^2\gamma$ and $\sin^2\gamma$, its contribution is
the same regardless of the polarity of the magnetic field (see Footnote 1). Treating the real
Zeemann pattern as an effective triplet ($J=1 \rightarrow J=0$), the functions $\phi_p$, $\phi_r$
and $\phi_b$ refer to the Voigt profiles for the $\Pi$ ($\Delta M=0$), blue ($\Delta M = -1$)
and red ($\Delta M = 1$) components of the Zeeman pattern, respectively. The former is centered at $\lambda_0$ 
(central laboratory wavelength), whereas the latter ones are shifted by an amount $\pm \lambda_B$ with respect
to the $\Pi$ component.

\begin{equation}
\lambda_B = \mathcal{C} g_{\rm eff} B \lambda_0^2
\end{equation}

\noindent where $C=4.67\times 10^{-13}$ [\AA\ Gauss]$^{-1}$, $B$ is the strength of the magnetic field  
(measured in Gauss), and $g_{\rm eff}$ is the effective Land\'e factor of the spectral line.

A number of remarks are in order. First of all, in the absence of a magnetic field (or if $g_{\rm eff}=0$), 
$\phi_p=\phi_r=\phi_b$ and therefore $\eta_I=1+\eta_0 \phi_p/2$. Also, it is interesting to see that 
the blue and red components are always present independently
of the orientation of the field. When the field is small enough the splitting is much smaller
that the Doppler width of the spectral line, $\lambda_B \ll \Delta \lambda_{\rm D}$,
 causing the spectral line to broaden. In the opposite case, $\lambda_B \gg \Delta \lambda_{\rm D}$, 
they appear as two separate spectral lines with half the strength of the spectral line in the absence of 
magnetic field. Finally, the central $\Pi$ component vanishes for magnetic fields
aligned with the observer ($\gamma=0$) causing the line core to desaturate.

\section{Effects on the magnetic field in the abundance of Fe, Si, C and O}

We now proceed to calculate synthetic Stokes $I$ profiles of the spectral lines
in Table 1. We use the numerical code and the atmospheric model described in Sect~2.1.
We consider a magnetic field with a varying strength and inclination: $B=[0,500]$ Gauss,
$\gamma=[0,90]$ deg. We consider also standard solar abundances: $\log\epsilon_{Fe}=7.45$ (Asplund et al. 2000),
$\log\epsilon_{Si}=7.51$ (Asplund 2000), $\log\epsilon_{C}=8.39$ (Asplund et al. 2005b)
and $\log\epsilon_{O}=8.66$ (Asplund et al. 2004). The resulting profiles are then fitted
 with different abundances but assuming that there is no magnetic 
field. For \ion{Fe}{1} we use $\log\epsilon_{Fe}=[7.30,7.80]$ in steps of 0.01. For \ion{Si}{1}
and \ion{O}{1} we consider  $\log\epsilon_{Si}=[7.48,7.54]$ and $\log\epsilon_{O}=[8.62,8.70]$,
respectively, both in steps of 0.002. Finally, for \ion{C}{1} we use  $\log\epsilon_{C}=[8.32,8.46]$
in steps of 0.005. 

From the comparison of the original profiles with magnetic field
and fixed abundance with those where the abundance varies and the magnetic field is neglected,
we obtain a $\chi^2-\log\epsilon$ curve for each spectral line. We therefore infer for each
line an optimum abundance, as the one that minimizes the $\chi^2$ curve. No other free parameters
are considered. An example of this process is presented 
in Figure 1 using the spectral line \ion{Fe}{1} 6136.994 \AA\, a magnetic field of $B=250$ Gauss and 
$\gamma=60^{\circ}$. The final inferred abundance is obtained as the mean of the best-fit abundances 
from all spectral lines of that atomic element.

\begin{center}
\includegraphics[width=7.5cm]{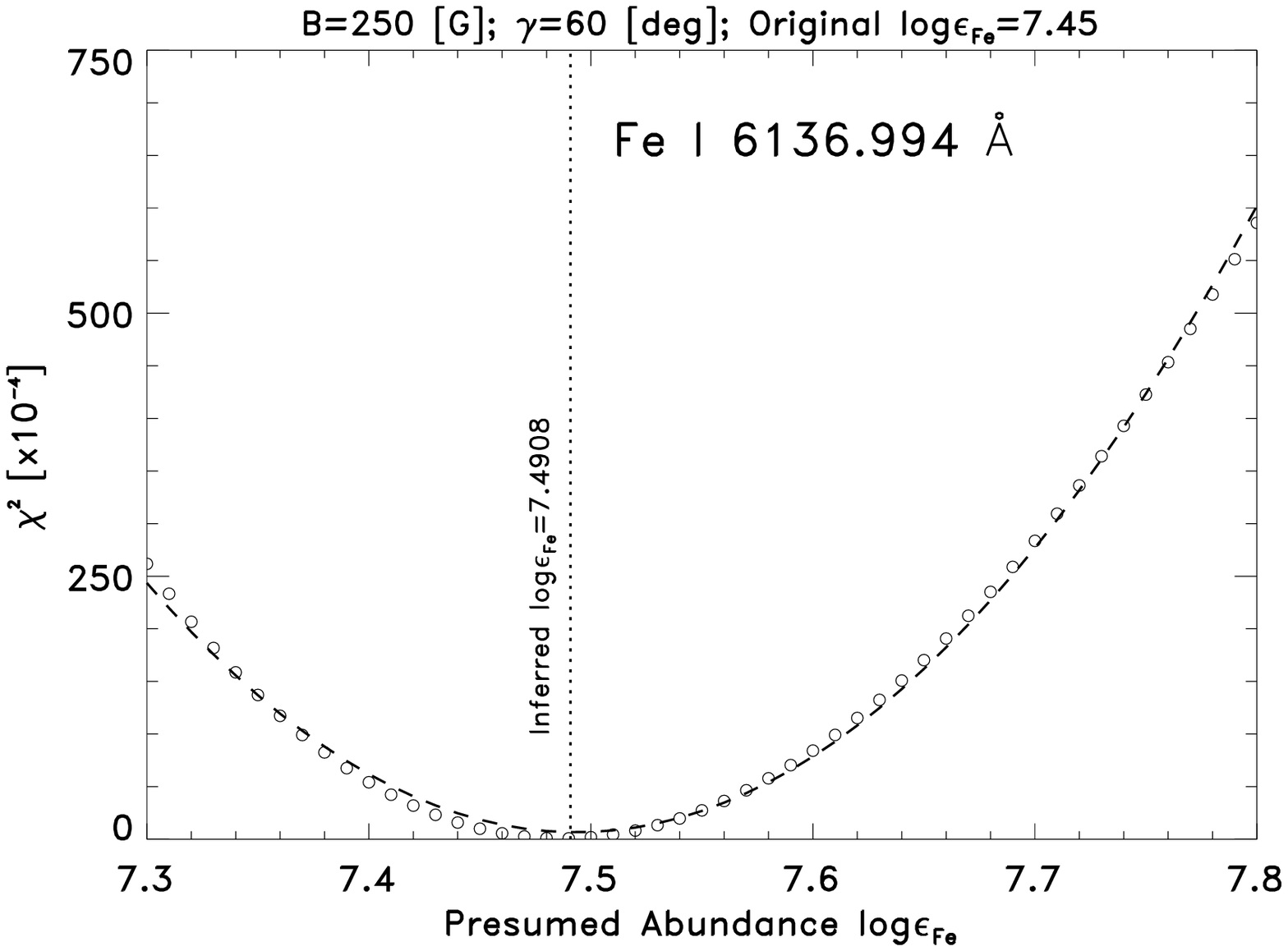}
\figcaption{$\chi^2$ between the intensity profile of \ion{Fe}{1} 6136.994 \AA\ with
magnetic field ($B=250$ Gauss; $\gamma=60^{\circ}$) and an iron abundance of $\log\epsilon_{Fe}=7.45$, 
and the intensity profile of the same spectral line without magnetic field, as a function of the abundance. 
The abundance that produces the best fit to the former profile is $\log\epsilon_{Fe}\simeq7.49$.}
\end{center}

Figure 2 presents the errors $\Delta\log\epsilon=\log\epsilon_{\rm fit}-\log\epsilon_{\rm real}$ 
(as a function of the magnetic field strength and inclination) introduced in the abundance of
 \ion{Fe}{1}, \ion{Si}{1}, \ion{C}{1} and \ion{O}{1}, when the magnetic field is not accounted for. As 
expected, the larger the field strength the larger the error. Neutral iron, 
with up to $\Delta\log\epsilon_{\rm Fe} \le 0.1$, presents the largest deviations. For the rest of considered 
elements the magnetic field seems to have only a marginal effect: $\Delta\log\epsilon_{\rm Si} \le 0.01$ and $\Delta\log\epsilon_{\rm C,O} 
\le 0.02$.

We also find that vertical fields underestimate the correct abundance, whereas the opposite happens for more inclined 
magnetic fields. This can be explained attending to Sect~2.3 (Equation 1). For vertical 
magnetic fields, $\gamma=0^{\circ}$, the $\Pi$ component of the Zeemann pattern
is absent, leading to a desaturation of the core intensity and therefore requiring a smaller abundance to fit 
the line profile. An example with \ion{Fe}{1} 5250.209 \AA\ is presented in Figure 3 (left panel). 
However, when the magnetic field is horizontal,  $\gamma=90^{\circ}$, the line mainly broadens and 
thus yieling a larger abundance. See example for \ion{Fe}{1} 6200.313 \AA\ in Figure 
3 (right panel).

\begin{center}
\includegraphics[width=7.5cm]{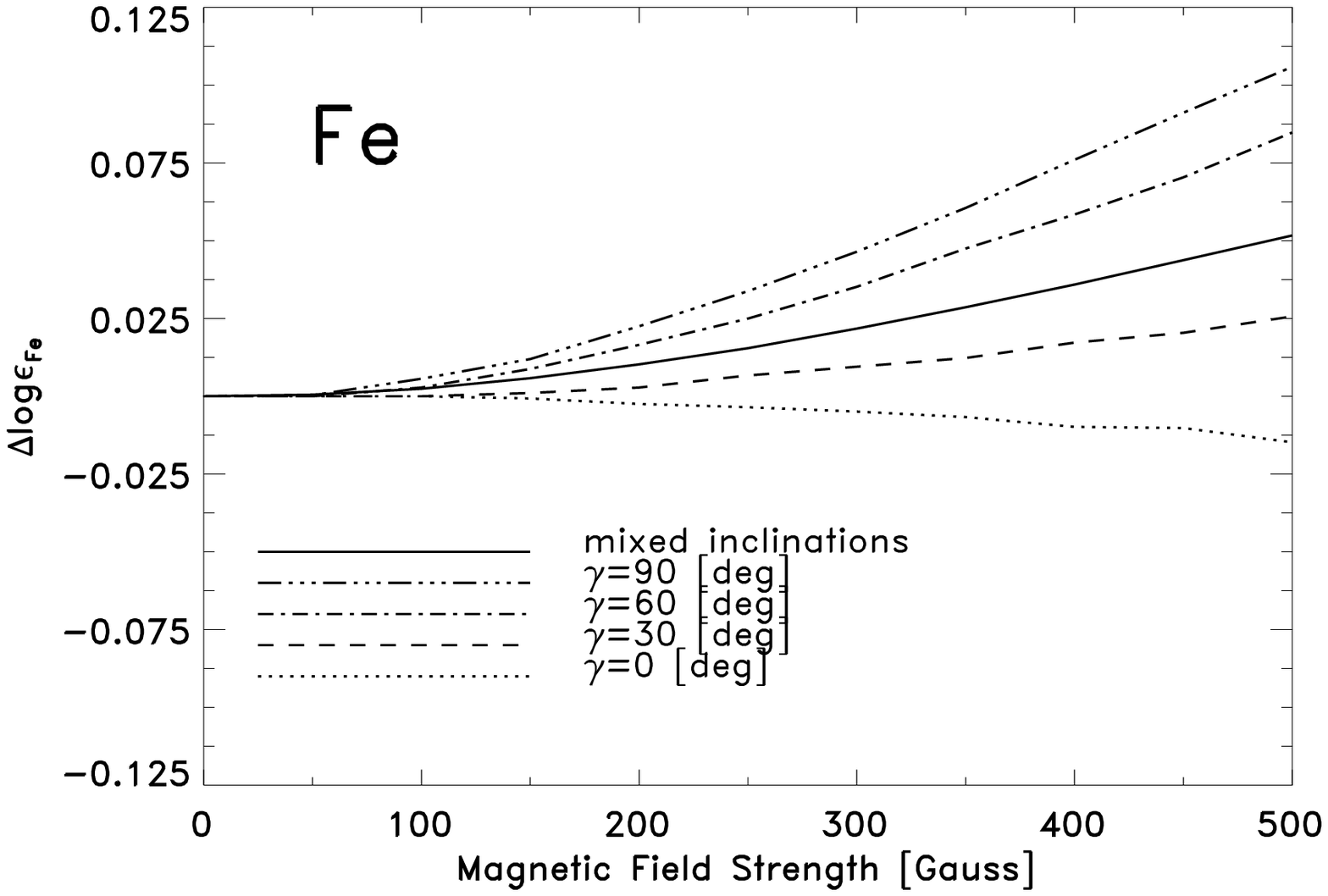}
\includegraphics[width=7.5cm]{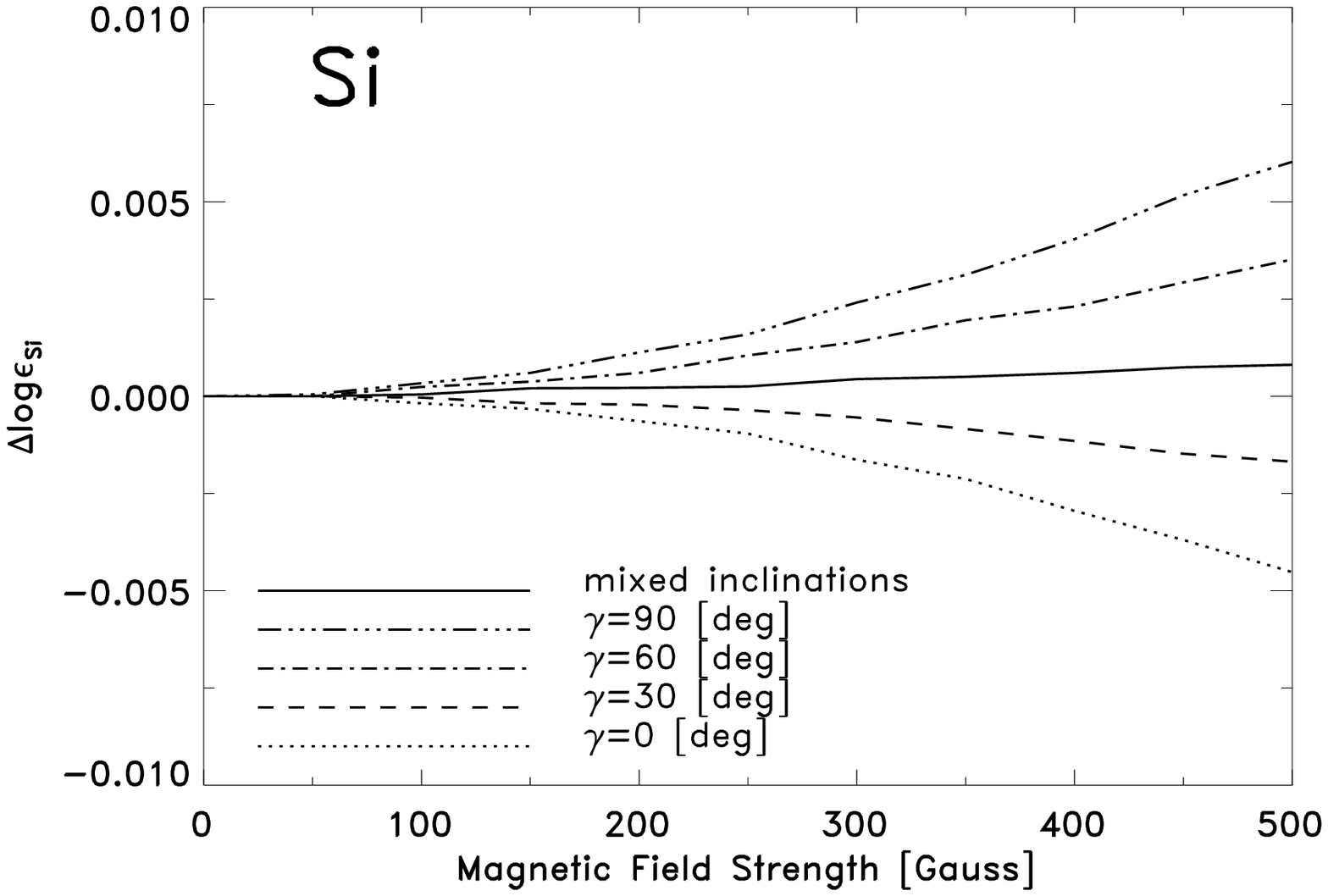}
\includegraphics[width=7.5cm]{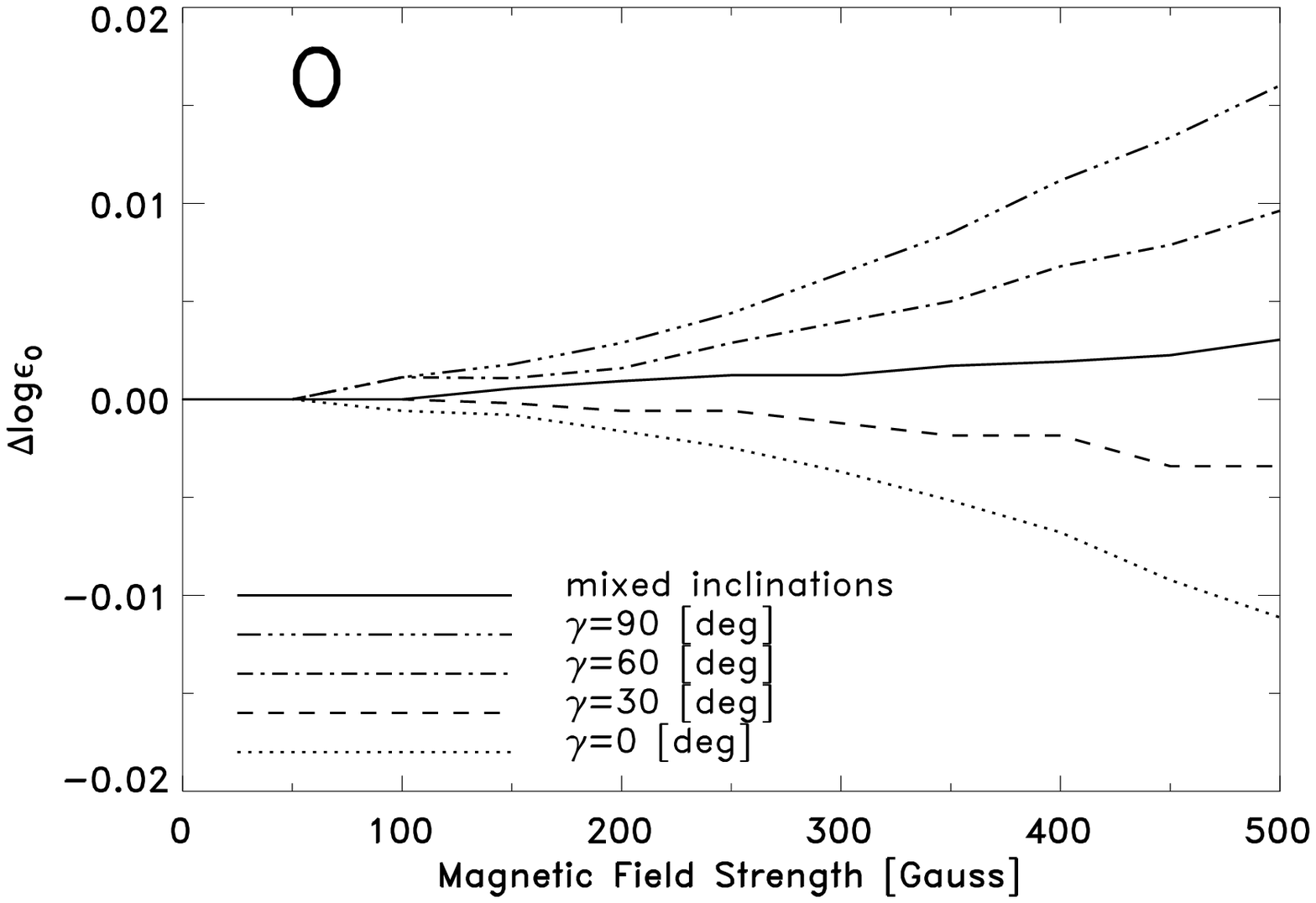}
\includegraphics[width=7.5cm]{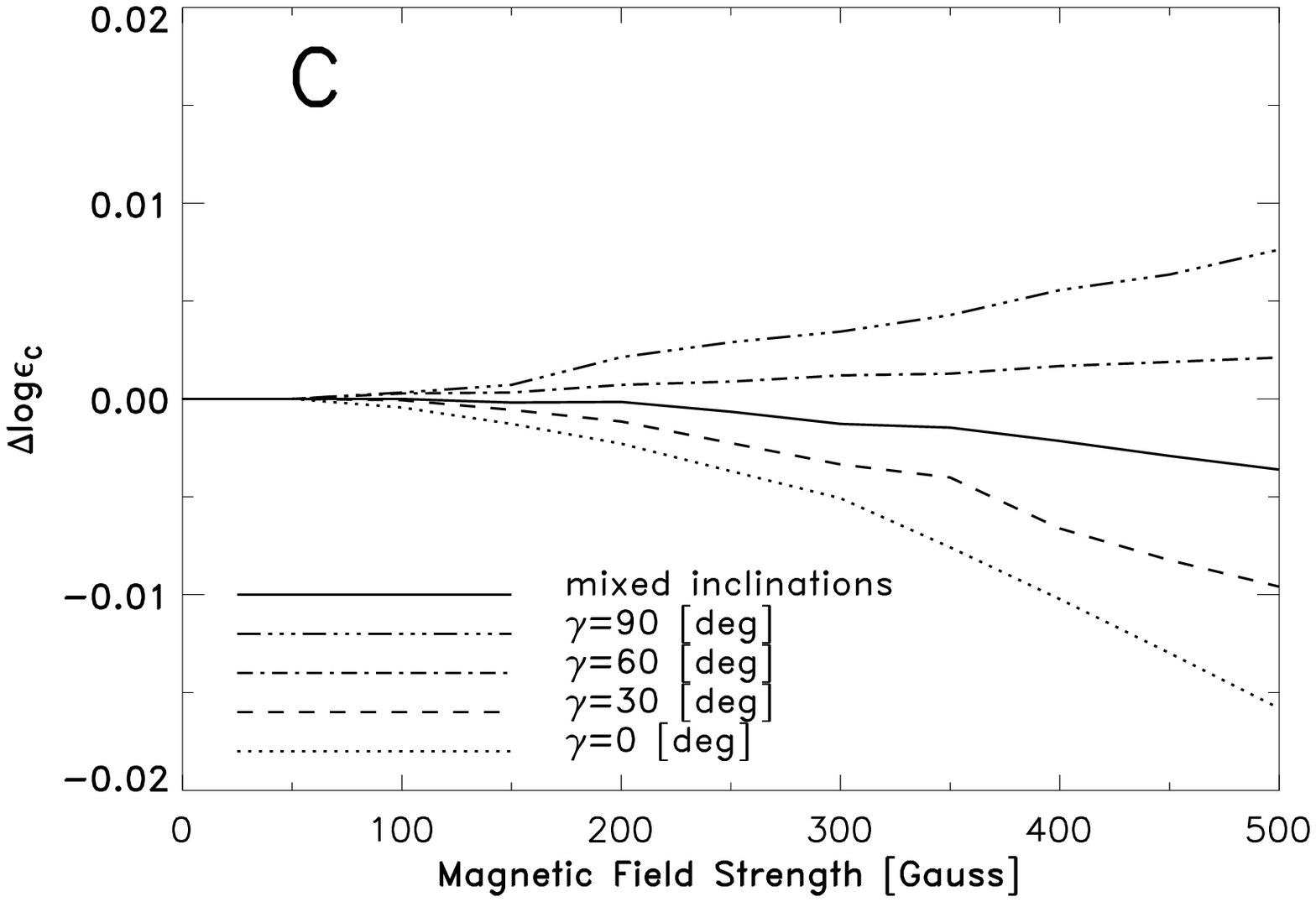}
\figcaption{Errors in the inferred abundances $\Delta\log\epsilon=\log\epsilon_{\rm fit}-\log\epsilon_{\rm real}$
when the role of the magnetic field is neglected, as a function of the magnetic field strength for 4 different
inclinations: $\gamma=0^{\circ}$ (dotted line; vertical magnetic field), $\gamma=30^{\circ}$ (dashed), $\gamma=60^{\circ}$
(dashed-dotted) and $\gamma=90^{\circ}$ (dashed-triple dotted; horizontal magnetic field). Solid lines indicate a mixture
of field inclinations, in which profiles obtained with the previous 4 inclinations are averaged before inferring an abundance.}
\end{center}

We must also consider that different regions of solar and stellar atmospheres can posses magnetic
fields with different inclinations. For instance, in the case of the solar granulation, where the 
field is mostly horizontal in the granules (upflowing gas), but is generally vertical in
intergranular lanes (downflowing gas). Since the inclination that matters is with respect to
the observer, this situation reverses as we move towards higher latitudes or towards the solar limb.
Many other low and intermediate-mass stars also posses an outer convective layer. In addition, strong magnetic concentrations 
also display a variety of inclinations: star-sunspots (umbra and penumbra), 
pores, network regions etc. Although they are normally avoided in the Sun, in the stellar case they certainly
contribute to the observed profiles.

Since vertical and horizontal fields seems to have opposite effects in the derived abundance it
is appropriate to study whether they can cancel each other out when a mixture of different 
inclinations is present. To study this effect we have carried out a similar experiment as the ones previously
presented, being now the difference that we average the intensity profiles obtained with 4 different inclinations,
$\gamma=0,30,60,90^{\circ}$ (with the same field strength) before inferring an abundance for each spectral line. 
In this case, the errors in the retrieved abundance are much smaller (see solid line in Figure 2), being only
perceptible for the case of neutral iron: $\Delta\log\epsilon_{\rm Fe} \le 0.02$.
This simulation is very simple in the sense that it does not include different temperature
stratifications, different velocity fields, etc. However, it helps to highlight that the aforementioned cancellation effect 
can indeed take place. This cancellation effect in Stokes $I$ due to vertical and horinzontal magnetic fields, is similar in a way
to the cancellation of circular polarization signals due to a mixture of magnetic fields pointing towards and away from
the observer.

\begin{center}
\includegraphics[width=7.5cm]{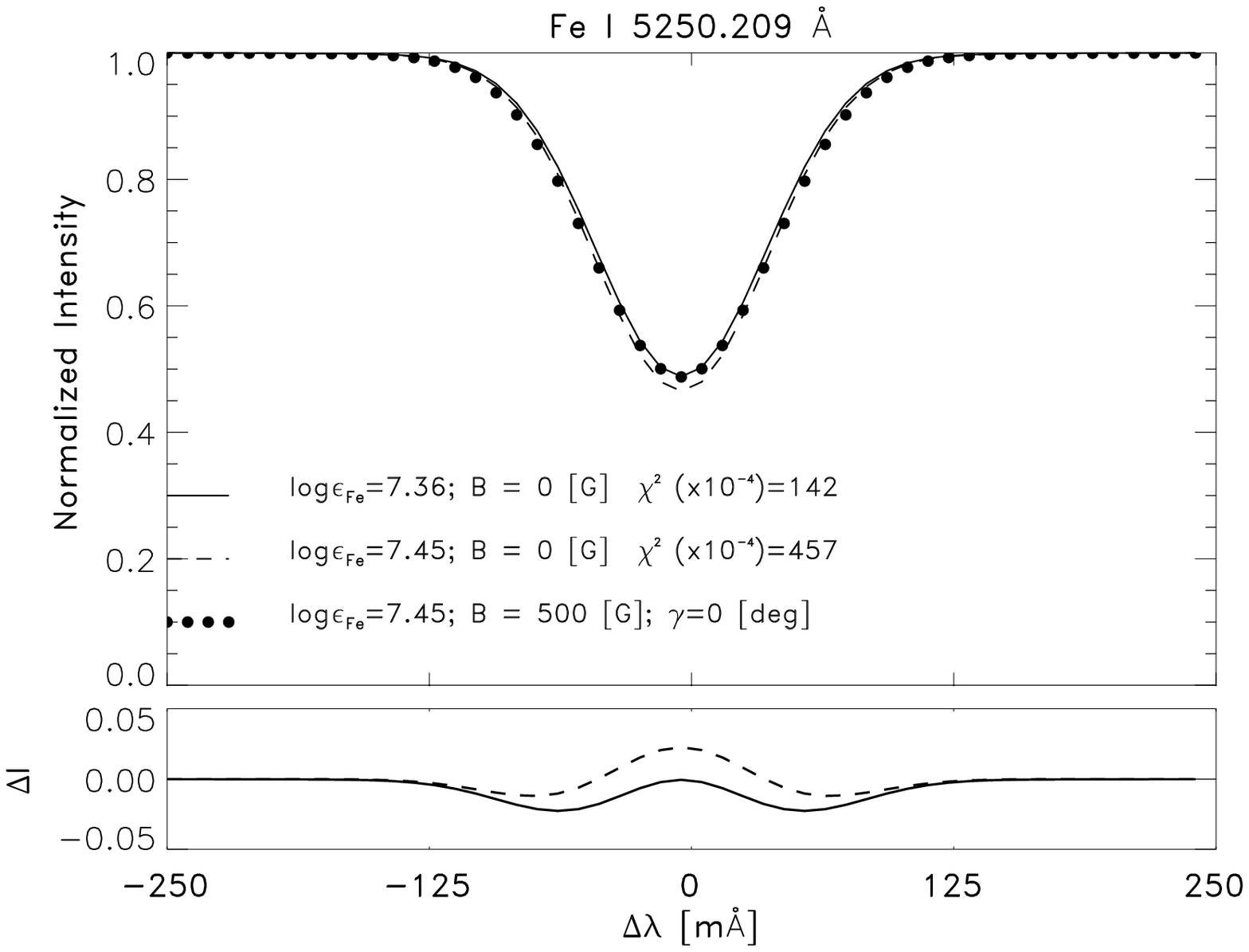}
\includegraphics[width=7.5cm]{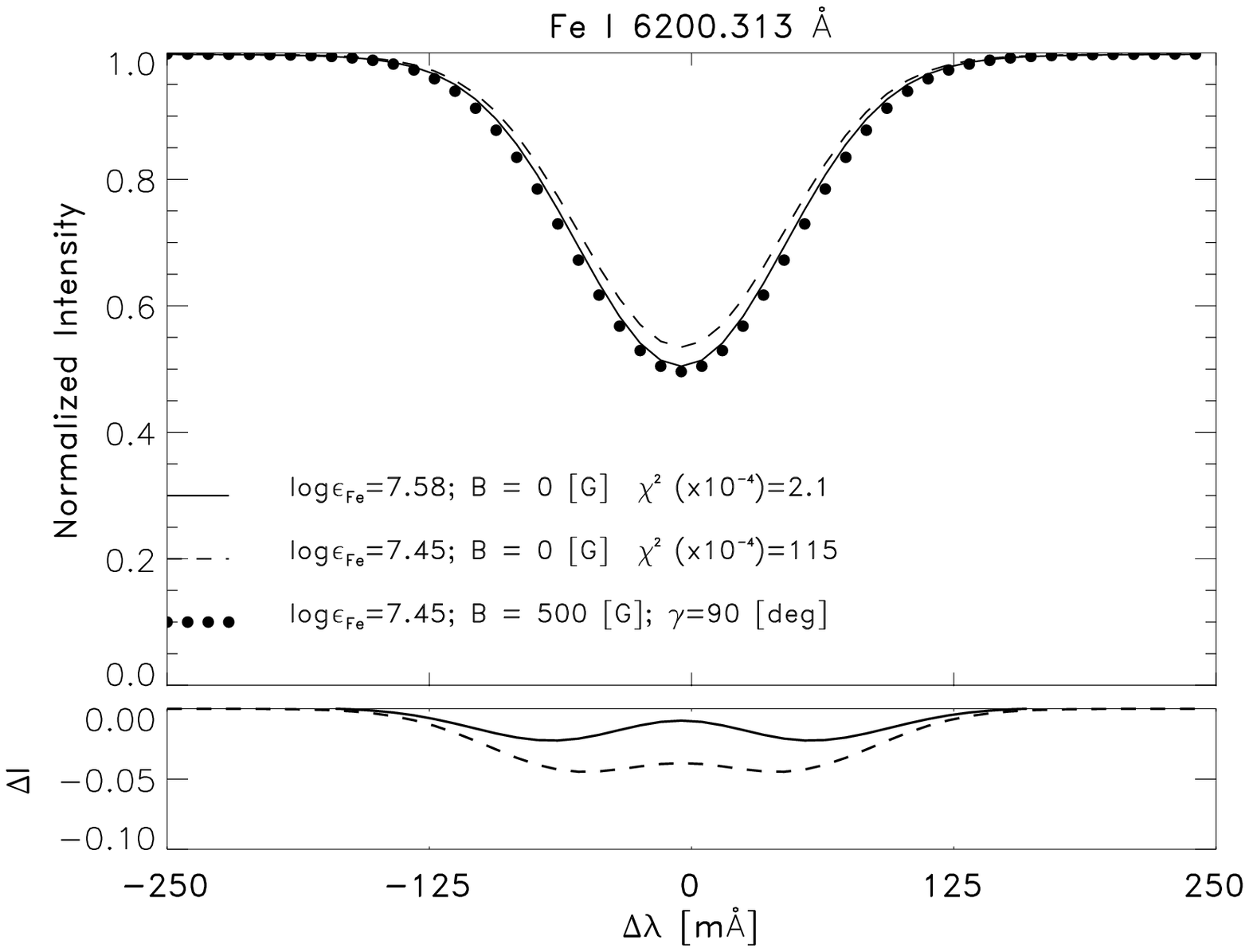}
\figcaption{Filled circles are intensity profiles obtained with the nominal iron abundance of $\log\epsilon_{Fe}=7.45$, 
a magnetic field strength of $B=500$ Gauss and inclinations of $\gamma=0^{\circ}$ (left panel) and $\gamma=90^{\circ}$
(right panel). Those profiles are then fitted without a magnetic field. When the original profile is affected
by a vertical magnetic field, the best fit (solid lines) is obtained by underestimating the original abundance. The 
opposite occurs for horizontal magnetic fields.}
\end{center}

From these results we conclude that the effect of neglecting the role of the magnetic field in abundance determinations 
strongly depends on the atomic specie and spectral line under consideration, as well as the strength and inclination of the magnetic field.
From the four atomic species considered, it appears to have a measurable effect in the case of \ion{Fe}{1} only. It could be 
marginally important for \ion{Si}{1}, \ion{C}{1} and \ion{O}{1} only if the magnetic field had a clearly preferred orientation. 
In stellar atmospheres, where we observe disk-integrated signals, this is unlikely the case. In addition, magnetic fields do not appear
as a plausible source of error in the current controversy of the solar carbon and oxygen abundance.

\section{A phenomenological model}

The different behavior seen in \ion{Fe}{1} as compared to the other three elements considered cannot
be understood in terms of the magnetic sensitivity of the lines used, as most
of them have similar Land\'e factors ranging from $g_{\rm eff}=1-2$. The source of these differences must be
therefore thermodynamic. In this section we will develop a tool to differentiate whether a given
spectral line of a particular atomic element is prone to yield unreliable abundances.
To that end we assume that the error in the inferred abundances, that appears as a consequence of 
neglecting the magnetic field, is directly proportional to the changes in the line profile in the
presence of a magnetic field:

\begin{equation}
\Delta \log \epsilon \propto \max \left\|\frac{\partial I}{\partial B}\right\|
\end{equation}

\noindent where we use maximum of the absolute value since the derivative
changes sign as a function of wavelength. Attending Equation 1, we will
model the intensity profile as a combination of three Gaussian, where
two of them are shifted by an amount $\pm \lambda_B$ with respect
to the central laboratory wavelength $\lambda_0$. The third Gaussian 
is centered at $\lambda_0$ and possesses twice the strength of the other two.

\begin{align}
I(\lambda) = 1 - \frac{[1-I_0] \sin^2\gamma}{2} e^{-\frac{(\lambda-\lambda_0)^2}{\Delta\lambda^2}}\\-\frac{[1-I_0](\cos^2\gamma+1)}{4} 
\left\{e^{-\frac{(\lambda-\lambda_0-\lambda_B)^2}{\Delta\lambda^2}} +e^{-\frac{(\lambda-\lambda_0+\lambda_B)^2}{\Delta\lambda^2}}\right\}\notag
\end{align}

\noindent where $I_0$ and $\Delta\lambda$ refer to the core intensity and the Doppler width of
the spectral line in the absence of magnetic field. That is,

\begin{equation}
\lim_{B \to 0}  I(\lambda) = 1 - (1-I_0) e^{-\frac{(\lambda-\lambda_0)^2}{\Delta\lambda^2}}
\end{equation}

Note that $\Delta\lambda$ is related to the HWHM of the Gaussian profile by a constant
factor: HWHM=$\sqrt{\log 2} \Delta\lambda \simeq 0.832\Delta\lambda$. Table 1 presents
the values of $I_0$ and $\Delta\lambda$ obtained from a Gaussian fit to the intensity profiles
in the absence of magnetic field for each spectral line. Finally, Equations (2)-(4) can be combined
to write:

\begin{align}
\Delta \log \epsilon \propto \max \bigg\| \frac{1-I_0}{2\Delta\lambda^2} (\cos^2\gamma+1)
 \mathcal{C} g_{\rm eff} \lambda_0^2 [(\lambda-\lambda_0 +\notag \\ + \lambda_B) e^{-\frac{(\lambda-\lambda_0+\lambda_B)^2}{\Delta\lambda^2}}-
(\lambda-\lambda_0-\lambda_B) e^{-\frac{(\lambda-\lambda_0-\lambda_B)^2}{\Delta\lambda^2}}] \bigg\|
\end{align}

This procedure is very similar to the one used in Cabrera Solana et al. (2005) but using Stokes
$I$ instead of the circular polarization, Stokes $V$. Equation 6 allows to predict the effect of ignoring the magnetic field
using: properties of the spectral line (regardless of the atomic specie) 
in the absence of a magnetic field ($I_0$, $\Delta\lambda$, $\lambda_0$), Land\'e factor $g_{\rm eff}$ 
(given by the electronic configurations) and the strength of 
the magnetic field (used to calculate $\lambda_B$). Note that the inclination of the magnetic field
does not play any role. The factor $[\cos^2\gamma+1]$ does not help to explain the fact that for
vertical fields $\Delta\log\epsilon<0$, but $\Delta\log\epsilon>0$ for horizontal fields. Therefore
it is more appropriate to write:

\begin{align}
\Delta \log \epsilon = \mathcal{A}(\gamma) \max \bigg\| \frac{1-I_0}{2\Delta\lambda^2}
 \mathcal{C} g_{\rm eff} \lambda_0^2 [(\lambda-\lambda_0+ \notag \\ + \lambda_B) e^{-\frac{(\lambda-\lambda_0+\lambda_B)^2}{\Delta\lambda^2}}-
(\lambda-\lambda_0-\lambda_B) e^{-\frac{(\lambda-\lambda_0-\lambda_B)^2}{\Delta\lambda^2}}] \bigg\|
\end{align}

\noindent where the calibration constant $\mathcal{A}(\gamma)$ can be determined for different
field inclinations using our results in Section 3. This is done in Figure 4, where we plot
the left-hand side term of Equation 6 (obtained from Section 3) versus the right hand-side of 
the same equation (evaluated using Table 1) for two limiting inclinations: $\gamma=0^{\circ}$ 
(upper panel) and $\gamma=90^{\circ}$ (lower panel) for all spectral lines. As it can be seen
the correlation is good enough to empirically justify, in first approximation, our assumption in Equation 3.
That is, the more a spectral line is affected by the magnetic field, the more unreliable the 
derived abundance will be if the magnetic field is not considered. In addition, all spectral lines
seem to follow a linear relation regardless of the atomic specie. The present model
explains why \ion{Fe}{1} is affected by the magnetic field more than
\ion{Si}{1}, \ion{C}{1} and \ion{O}{1}. As already mentioned, this is not due to the different Land\'e factors, but rather due
to their different sensitivity to thermodynamic parameters: $I_0$ and $\Delta\lambda$. Narrower
and deeper spectral lines ($\Delta\lambda$ and $I_0$ small) such as the ones from
\ion{Fe}{1}, are more sensitive to the magnetic field than weak and broad spectral lines.

It is also interesting to notice that the selected \ion{C}{1} lines are broader and weaker than those of 
\ion{Si}{1} and \ion{O}{1}. Despite this, carbon is affected by the magnetic field as much as the other two 
(see Fig.~2). The reason is that the employed \ion{C}{1} lines are located at larger wavelengths, $\lambda_0^{C} > \lambda_0^{Si,O}$,
and therefore the Zeeman splitting is larger (Eq.~2 and 7).

Since we use the same temperature stratification, the different 
$I_0$'s and $\Delta\lambda$'s among the selected spectral lines can only be due to differences in  
the excitation potential of the lower level $\chi_l$, transition probability
$\log gf$, and collisional broadening parameters $\alpha$ and $\sigma$.
In first approximation, the former two would be related to $I_0$ whereas the latter two determine 
$\Delta\lambda$.

The utility of the procedure described here lies in the fact that whenever another spectral line
is considered (regardless of the atomic element), we can use this method to evaluate Equation 7, and therefore
calculate its approximate position in Figure 4. This will give us an idea of the error introduced
by ignoring the effects of the magnetic field, even if a numerical code that solves the radiative transfer
equation in the presence of a magnetic field is not available.

\begin{center}
\includegraphics[width=9cm]{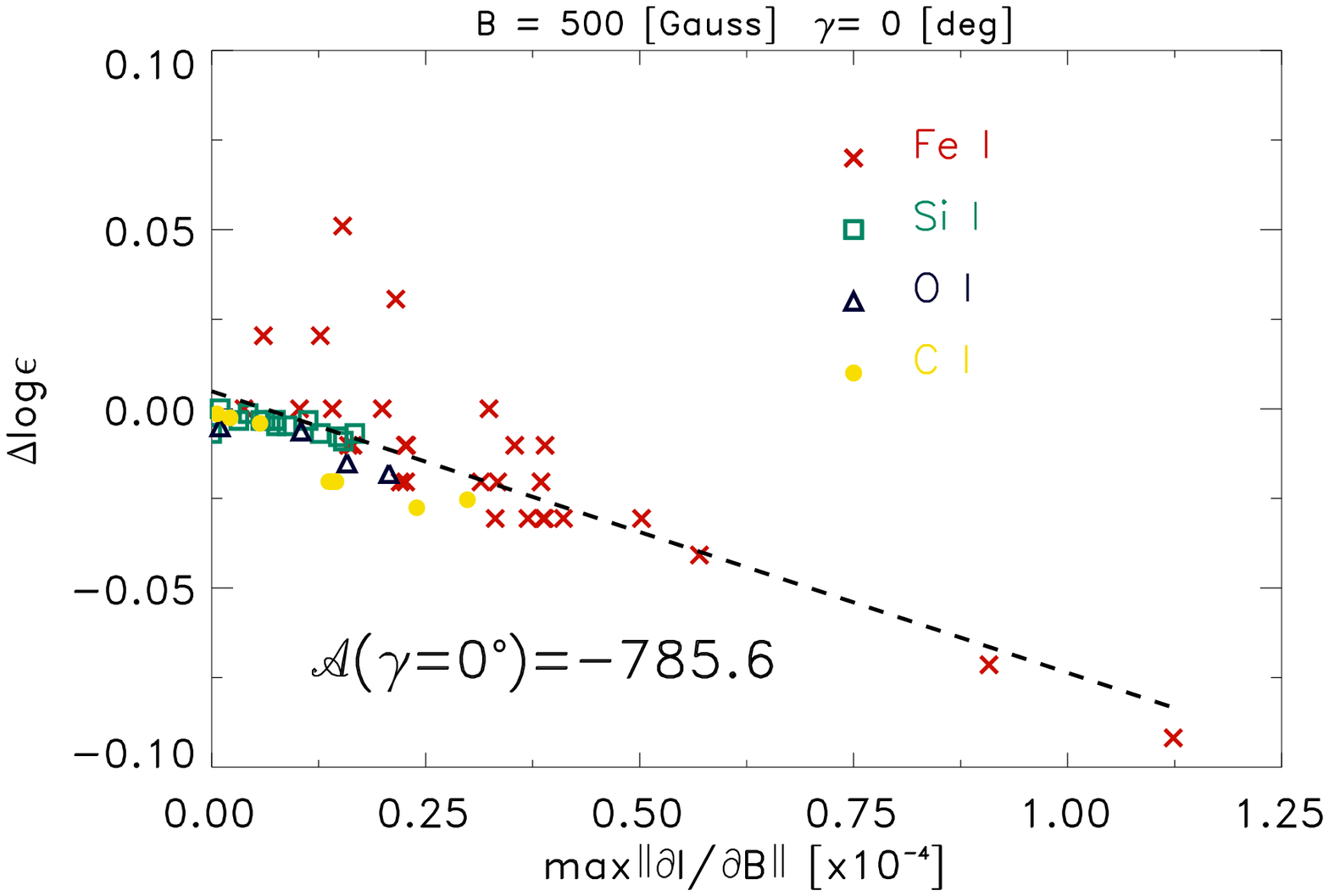}
\includegraphics[width=9cm]{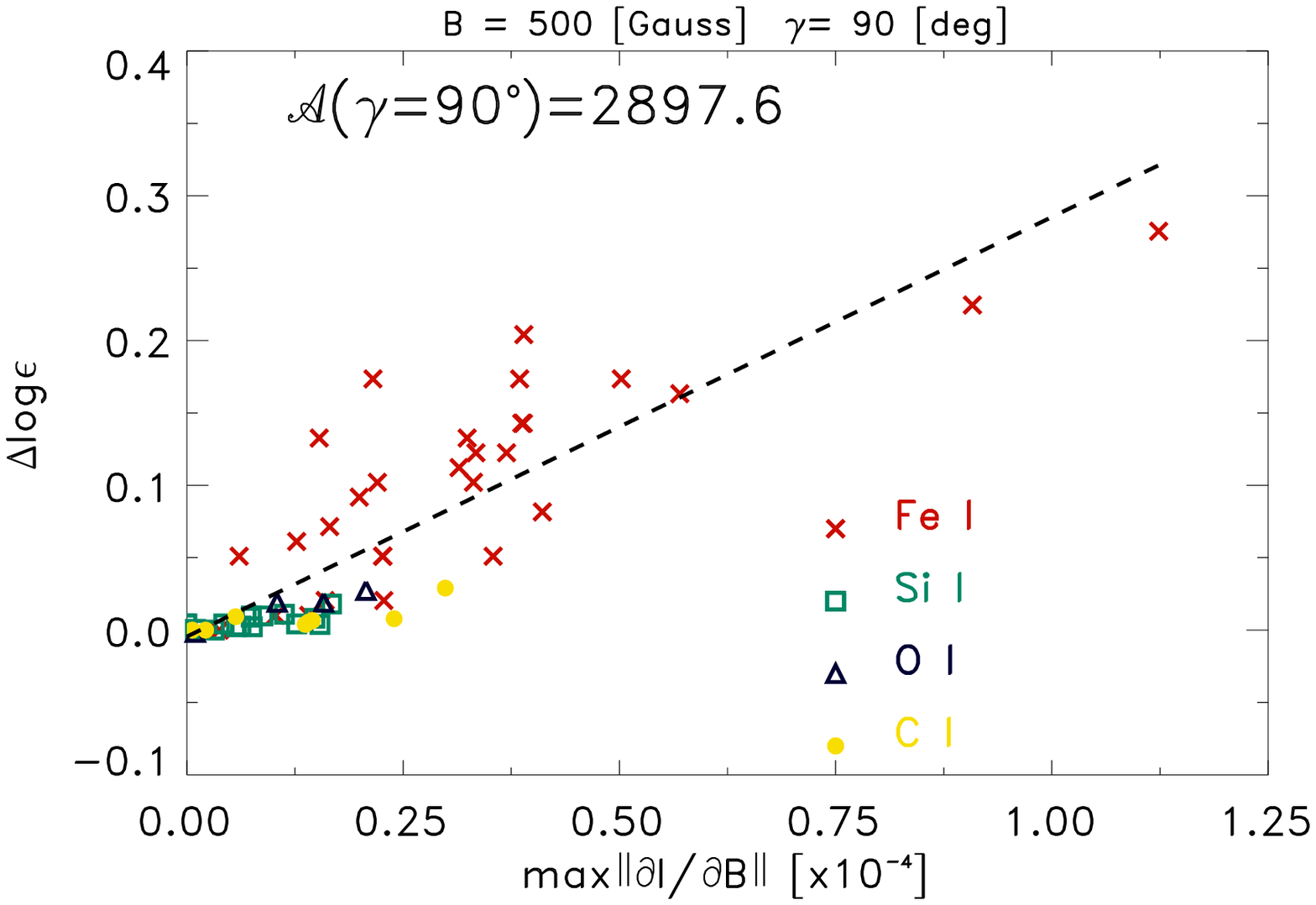}
\figcaption{Error in the determination of element abundances when using individual spectral lines versus 
the derivative of its intensity profile with respect to the magnetic field. Red crosses
show the 29 \ion{Fe}{1} lines from Table 1. Green squares indicate \ion{Si}{1}. \ion{O}{1} is denoted
by blue triangles. Yellow circles show \ion{C}{1} lines. A linear fit to the data points is
indicated by the black dashed line. The slope of the fitted line is indicated by $\mathcal{A}(\gamma)$.
{\it Top panel}: vertical magnetic field. {\it Bottom panel}: horizontal magnetic field.}
\end{center}

\section{Convective vs Magnetic broadening}

In our simple 1D model the convective broadening is introduced through the macro and
microturbulent velocities, whereas in a more realistic 3D modeling this broadening
naturally occurs when adding the profiles emerging from different regions (i.e.:
granules and intergranules) in the solar atmosphere. At first glance it seems plausible
for this convective broadening to mask the magnetic one, thus making the contribution 
of the magnetic fields even more negligible. However, we must take into account that the individual
intensity profiles emerging from different regions of the solar atmospheres are already affected by the
magnetic field. Therefore, when all intensity profiles are added up to produce the corresponding
convective broadening, the fingerprints of magnetic field will still be visible.

To prove this statement we have carried out a simple simulation of the process previously described.
To that end we synthesize intensity profiles of \ion{Fe}{1} 5247.05 \AA\ (g$_{\rm eff}$=2) 
without magnetic field and no net line-of-sight velocity, but using a macroturbulent velocity of 2 km s$^{-1}$. 
The resulting profile is  plotted in Figure 5 ($I_1$, thick dashed line). We then repeat this synthesis
 but now including a vertical magnetic field, $\gamma=0^{\circ}$, with a strength of 1000 Gauss. 
The resulting profile is plotted in Figure 1 ($I_2$, thick solid line).

Next we perform a similar synthesis, where: a) we set the macroturbulent velocity to zero;
b) we include a vertical field of 1000 Gauss and, c) we use a net line-of-sight velocity that
changes from $-$4 km s$^{-1}$ to 4 km s$^{-1}$. Each of the emerging profiles is indicated
by one of the thin dashed lines in Fig.~1 (labeled as $I_k$). With this we try to simulate the effect of having 
different structures affected by convective velocity fields. We then add all those profiles
using a weighting function with a Gaussian shape in order to mimic the macroturbulence. 
The resulting profile is indicated in Fig.~1 as $I_3$ (squares).

\begin{eqnarray}
I_3 = \sum w_k I_k \\
w_k = \frac{v_{\rm mac}}{2} \sqrt{\frac{\pi}{\log2}} e^{-4\log2 \frac{v_{\rm los,k}^2}{v_{\rm mac}^2}}
\end{eqnarray}

where the FWHM of the Gaussian weighting function is equivalent to a macroturbulent velocity of 2 km s$^{-1}$. 
As it can be seen, $I_3 = I_2$. Therefore, the broadening due to convective 
motions does not hide the broadening produced by magnetic
fields. Our simulations do not consider that this magnetic field is probably different in granules 
(upflows) and intergranules (downflows). We also use the same temperature stratification
in all cases. Considering a more realistic case (using results from 3D MHD simulations)
might yield slightly different results, but the same basic idea will apply also in that case.
In this example we have used a relatively strong magnetic field (1000 Gauss) to facilitate
a visual comparisons in Fig.~5, but it is worth mentioning that repeating the experiment with smaller fields leads 
to the same result.

\begin{center}
\includegraphics[width=8.5cm]{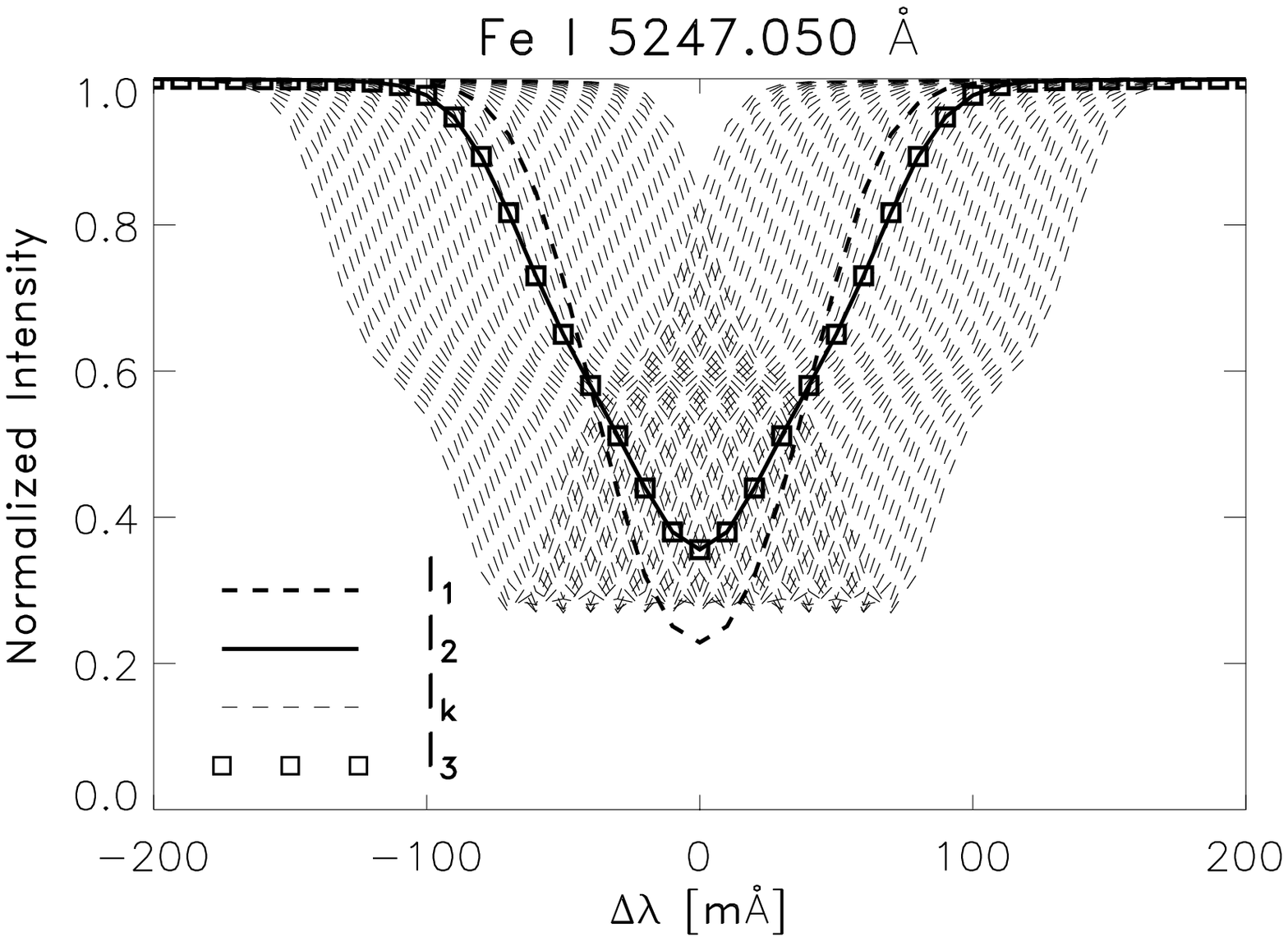}
\figcaption{Simulated intensity profiles of \ion{Fe}{1} 5257.050 \AA\. Thick dashed line, $I_1$
was obtained using a macroturbulent velocity of $v_{\rm mac}=2$ km s$^{-1}$. In addition to this,
we include a magnetic field with: $B=1000$ Gauss and $\gamma=0^{\circ}$, thus obtaining $I_2$ (thick solid).
Individual thin dashed lines, $I_k$, were obtained with $v_{\rm mac}=0$,  $B=1000$ Gauss and $\gamma=0^{\circ}$
and a varying $v_{\rm los}$ from $-$4 km s$^{-1}$ to 4 km s$^{-1}$ to simulate a convective velocity field. All
$I_k$ are added up into $I_4$ according to Eq.~8 and 9.}
\end{center}

To provide further support to our argument we have recalculated the effects of
the magnetic fields in the abundance (as done in section 3) but using a smaller
macroturbulence: $v_{\rm mac}=1$ km s$^{-1}$. The resulting plots are identical to
those in Fig.~2. This result might seem at odds with Eq.~7, since the new
macroturbulent velocity changes the values of $\Delta\lambda$ and $I_0$.
However, if we repeat the calculation of the calibration curves in Fig.~4 no
differences are observed in $\Delta\log\epsilon$ as compared to the case
with $v_{\rm mac}=2$ km s$^{-1}$. In order for this to happen, the only possibility
is that the horizontal axis changes. This results in a different calibration constant
$\mathcal{A}(\gamma)$ but equal errors in the abundance.

\section{Conclusions}

Ignoring the broadening caused by the magnetic field when fitting the intensity profiles of spectral lines,
may lead to an erroneous determination of atomic abundances in the Sun and other magnetically active stars.
Although there have been previous works where the magnetic field has been considered (Kochukhov et al. 2004;
Socas-Navarro \& Norton 2007), to our knowledge this is the first systematic study aiming at quantifying the
role of the magnetic field. Our results indicate that \ion{Fe}{1} lines are more affected than lines from elements
like \ion{Si}{1}, \ion{C}{1} or \ion{O}{1}. We have shown that vertical magnetic fields lead to
an underestimation of the real abundance, while horizontal fields tend to overestimate it. In a more real
situation, where the spectral lines would receive contributions from magnetic fields with different inclinations,
these two effect are likely to cancel each other, making the contribution of the magnetic field almost negligible with
the exemption of perhaps \ion{Fe}{1}. It is important to mention that the degeneracy between magnetic fields
and abundances occurs only for small fields, since for very large magnetic fields, the spectral line
is fully split in its different Zeemann components (see for example Nesvacil et al. 2004).
We have also developed a phenomenological model that can used to determine if a particular
spectral line is suitable for abundances studies that do not consider the effect of the magnetic field.

Our analysis 1D LTE analysis neglects the effect of having variations in
temperature stratifications and/or convective velocities fields. Therefore,
our results apply only if all those others possible sources of errors
can be eliminated. If that was not the case, the error introduced by the
magnetic field would be of second importance. This situation is highlighted
by the work of Socas-Navarro \& Norton (2007), who in spite of consistently
considering the magnetic field, obtained strong discrepancies between
the inferred abundances in quiet solar regions (e.g.: granulation) and magnetic regions (e.g.: pores).
The source of those discrepancies is therefore to be ascribed to small ($< 100$ K) errors
in the temperature stratification.

It would be ideal to repeat this work using realistic 3D MHD simulations (using
several initial magnetic fluxes) in order to model more realistically the different 
temperatures, velocities, field strength and inclinations present in solar and stellar photospheres.
However, our work does provide a first hint that magnetic fields are an unlike source
of large errors in abundance determinations, unless a very particular spectral line
or a very particular magnetic configuration is present.

\acknowledgements{I wish to thank  Eberhard Wiehr for bringing
up the issue, during a talk I gave at the University of G\"ottingen in 2003, 
of using only spectral lines with zero Land\'e factor in abundance determinations. 
That warning led to consider the effects of the magnetic field whenever 
those were not available. Thanks also to Carlos Allende Prieto and an anonymous
referee for important suggestions and comments.}

\begin{onecolumn}
\begin{center}
\tabcaption{Summary of employed spectral lines; temperature
parameter $\alpha$ and cross section for collisions with neutral hydrogen 
(in units of Bohr's radius $a_0$) from ABO theory (Barklem et al. 2000);
Land\'e factor g$_{eff}$ is calculated under LS approximation from the
electronic configurations of the upper and lower levels.}
\begin{tabular}{ccccccccccc}
\tableline
Atom & $\lambda_0$ [\AA] & $\chi_l$ [eV] & $\log(gf)$ & $\alpha$ & $\sigma/a_0^2$
& $g_{eff}$ & I$_0$ & $\Delta\lambda$ [m\AA] & Lower Level & Upper Level\\
\tableline

\ion{Fe}{1} & 4389.245 & 0.052 & $-$4.583 & 0.249 &  217 & 1.50  & 0.338 & 52.30  & 5D3 & 7F2\\
\ion{Fe}{1} & 5247.050 & 0.087 & $-$4.946 & 0.253 &  206 & 2.00  & 0.464 & 59.36 & 5D2 & 7D3\\
\ion{Fe}{1} & 5250.209 & 0.121 & $-$4.938 & 0.253 &  207 & 3.00  & 0.471 & 59.21 & 5D0 & 7D1\\
\ion{Fe}{1} & 5412.798 & 4.434 & $-$1.761 & 0.280 &  971 & 0.97  & 0.845 & 60.04 & 3G4 & 5H4\\
\ion{Fe}{1} & 5525.544 & 4.230 & $-$1.084 & 0.238 &  748 & 1.50  & 0.559 & 66.65 & 5D0 & 5D1\\
\ion{Fe}{1} & 5701.544 & 2.559 & $-$2.216 & 0.237 &  361 & 1.12  & 0.439 & 70.38 & 3F4 & 3D3\\
\ion{Fe}{1} & 5784.658 & 3.396 & $-$2.530 & 0.244 &  796 & 1.87  & 0.781 & 64.04 & 5F3 & 5D4\\
\ion{Fe}{1} & 5956.694 & 0.859 & $-$4.605 & 0.252 &  227 & 0.70  & 0.621 & 64.91 & 5F5 & 7P4\\
\ion{Fe}{1} & 6082.710 & 2.223 & $-$3.573 & 0.271 &  306 & 2.00  & 0.745 & 65.48 & 5P1 & 3P1\\
\ion{Fe}{1} & 6136.994 & 2.198 & $-$2.950 & 0.265 &  280 & 2.00  & 0.545 & 70.27 & 5P2 & 5D1\\
\ion{Fe}{1} & 6151.618 & 2.176 & $-$3.299 & 0.263 &  277 & 1.83  & 0.639 & 67.74 & 5P3 & 5D2\\
\ion{Fe}{1} & 6173.335 & 2.223 & $-$2.880 & 0.266 &  281 & 2.50  & 0.535 & 71.16 & 5P1 & 5D0\\
\ion{Fe}{1} & 6200.313 & 2.608 & $-$2.437 & 0.235 &  350 & 1.50  & 0.522 & 73.08 & 3F2 & 3F3\\
\ion{Fe}{1} & 6219.280 & 2.198 & $-$2.433 & 0.264 &  278 & 1.60  & 0.436 & 77.74 & 5P2 & 5D2\\
\ion{Fe}{1} & 6240.646 & 2.223 & $-$3.230 & 0.272 &  301 & 1.00  & 0.634 & 68.97 & 5P1 & 3P2\\
\ion{Fe}{1} & 6265.133 & 2.176 & $-$2.550 & 0.261 &  274 & 1.58  & 0.456 & 76.98 & 5P3 & 5D3\\
\ion{Fe}{1} & 6271.278 & 3.332 & $-$2.703 & 0.247 &  720 & 1.50  & 0.821 & 69.31 & 5F5 & 7D5\\
\ion{Fe}{1} & 6280.618 & 0.859 & $-$4.387 & 0.253 &  223 & 1.45  & 0.565 & 70.22 & 5F5 & 7F5\\
\ion{Fe}{1} & 6297.793 & 2.223 & $-$2.740 & 0.264 &  278 & 1.00  & 0.507 & 74.31 & 5P1 & 5D2\\
\ion{Fe}{1} & 6322.685 & 2.588 & $-$2.426 & 0.238 &  345 & 1.50  & 0.520 & 75.02 & 3F3 & 3F4\\
\ion{Fe}{1} & 6481.870 & 2.279 & $-$2.984 & 0.243 &  308 & 1.50  & 0.587 & 73.46 & 3P2 & 5D2\\
\ion{Fe}{1} & 6498.939 & 0.958 & $-$4.699 & 0.253 &  226 & 1.37  & 0.698 & 69.81 & 5F3 & 7F3\\
\ion{Fe}{1} & 6581.210 & 1.485 & $-$4.680 & 0.245 &  254 & 1.30  & 0.859 & 68.69 & 3F4 & 5F4\\
\ion{Fe}{1} & 6593.870 & 2.433 & $-$2.422 & 0.247 &  321 & 1.15  & 0.496 & 79.91 & 3H5 & 5G5\\
\ion{Fe}{1} & 6609.110 & 2.559 & $-$2.692 & 0.245 &  335 & 1.15  & 0.589 & 75.30 & 3F4 & 3G4\\
\ion{Fe}{1} & 6750.152 & 2.424 & $-$2.621 & 0.241 &  335 & 1.50  & 0.542 & 79.42 & 3P1 & 3P1\\
\ion{Fe}{1} & 6945.205 & 2.424 & $-$2.482 & 0.243 &  331 & 1.50  & 0.519 & 84.01 & 3P1 & 3P2\\
\ion{Fe}{1} & 6978.851 & 2.484 & $-$2.500 & 0.241 &  337 & 1.50  & 0.536 & 83.43 & 3P0 & 3P1\\
\ion{Fe}{1} & 7723.208 & 2.279 & $-$3.617 & 0.242 &  304 & 1.17  & 0.794 & 82.56 & 3P2 & 3D3\\
\ion{Si}{1} & 5645.613 & 4.93  & $-$2.04  & 0.223 & 1791 & 1.75  & 0.748 & 74.38 & 3P1 & 3S1\\
\ion{Si}{1} & 5665.555 & 4.92  & $-$1.94  & 0.222 & 1772 & 0.00  & 0.708 & 75.76 & 3P0 & 3P0\\
\ion{Si}{1} & 5684.484 & 4.95  & $-$1.55  & 0.221 & 1797 & 1.25  & 0.570 & 82.41 & 3P2 & 3S1\\
\ion{Si}{1} & 5690.425 & 4.93  & $-$1.77  & 0.222 & 1772 & 1.50  & 0.646 & 78.42 & 3P1 & 3P1\\
\ion{Si}{1} & 5701.104 & 4.93  & $-$1.95  & 0.222 & 1767 & 1.50  & 0.716 & 76.06 & 3P1 & 3P0\\
\ion{Si}{1} & 5708.400 & 4.95  & $-$1.37  & 0.222 & 1787 & 1.50  & 0.509 & 87.37 & 3P2 & 3P2\\
\ion{Si}{1} & 5772.146 & 5.08  & $-$1.65  & 0.207 & 2036 & 1.00  & 0.660 & 81.16 & 1P1 & 1S0\\
\ion{Si}{1} & 5780.384 & 4.92  & $-$2.25  & 0.228 & 1691 & 0.50  & 0.819 & 74.28 & 3P0 & 3D1\\
\ion{Si}{1} & 5793.073 & 4.93  & $-$1.96  & 0.228 & 1703 & 1.00  & 0.723 & 77.01 & 3P1 & 3D2\\
\ion{Si}{1} & 5797.856 & 4.95  & $-$1.95  & 0.223 & 1755 & 1.16  & 0.727 & 77.27 & 3P2 & 3D3\\
\ion{Si}{1} & 5948.541 & 5.08  & $-$1.13  & 0.222 & 1845 & 1.00  & 0.491 & 96.52 & 1P1 & 1D2\\
\ion{Si}{1} & 7680.266 & 5.86  & $-$0.59  & 0.495 & 2106 & 1.00  & 0.631 & 126.0 & 1P1 & 1D2\\
\ion{Si}{1} & 7918.384 & 5.95  & $-$0.51  & 0.232 & 2933 & 0.75  & 0.643 & 141.8 & 3D1 & 3F2\\
\ion{Si}{1} & 7932.348 & 5.96  & $-$0.37  & 0.235 & 2985 & 1.00  & 0.610 & 150.3 & 3D2 & 3F3\\
\ion{Si}{1} & 7970.307 & 5.96  & $-$1.37  & 0.232 & 2927 & 0.92  & 0.883 & 117.0 & 3D2 & 3F2\\
\ion{C}{1}  & 7111.469 & 8.640 & $-$1.074 & 0.313 & 1842    &  0.75 & 0.941 & 121.1 & 3D1    &  3F2\\
\ion{C}{1}  & 7113.179 & 8.647 & $-$0.762 & 0.314 & 1857    &  1.12 & 0.903 & 125.4 & 3D3    &  3F4\\
\ion{C}{1}  & 9603.036 & 7.480 & $-$0.895 & 0.236 &  561    &  2.00 & 0.752 & 175.4 & 3P0    &  3S1\\
\ion{C}{1}  & 10753.976& 7.488 & $-$1.598 & 0.238 &  532    &  2.00 & 0.875 & 174.5 & 3P2    &  3D1\\
\ion{C}{1}  & 11777.546& 8.643 & $-$0.490 & 0.271 &  746    &  0.92 & 0.854 & 214.1 & 3D2    &  3F2\\
\ion{C}{1}  & 12549.493& 8.847 & $-$0.545 & 0.293 &  863    &  1.50 & 0.880 & 228.0 & 3P0    &  3P1\\
\ion{C}{1}  & 12562.124& 8.848 & $-$0.504 & 0.293 &  863    &  1.50 & 0.874 & 230.6 & 3P1    &  3P0\\
\ion{C}{1}  & 12569.042& 8.848 & $-$0.586 & 0.293 &  863    &  1.50 & 0.885 & 226.3 & 3P1    &  3P1\\
\ion{C}{1}  & 12581.585& 8.848 & $-$0.509 & 0.293 &  862    &  1.50 & 0.874 & 230.9 & 3P1    &  3P2\\
\ion{O}{1}  & 6158.176 & 10.74 & $-$0.30 & 0.322 &  1915 &  1.58  & 0.976 &  98.30 & 5P3 & 5D3\\
\ion{O}{1}  & 7771.944 & 9.15  &  0.37   & 0.234 &   452 &  1.33  & 0.734 & 134.04 & 5S2 & 5P3\\    
\ion{O}{1}  & 7774.166 & 9.15  &  0.22   & 0.234 &   452 &  1.92  & 0.761 & 129.76 & 5S2 & 5P2\\    
\ion{O}{1}  & 7775.388 & 9.15  &  0.00   & 0.234 &   452 &  1.75  & 0.800 & 124.15 & 5S2 & 5P1\\
\tableline
\end{tabular}
\end{center}
\end{onecolumn}


\end{document}